%
%
%
%
%
%
%
\magnification=1200
                  \baselineskip=23pt
\footline{\hfill-- \folio\ --\hfill}
\font\boldit=cmbxti10
\textfont9=\boldit
\def\bmit {\fam9 \boldit}
\def\linespace#1{\vskip#1 \baselineskip}
\font\fontB=cmr10 scaled \magstep3

\centerline{\bf \fontB  Continued-fraction expansion of}
\centerline{\bf \fontB  eigenvalues of generalized evolution operators}
\centerline{\bf \fontB  in terms of periodic orbits}
\linespace1

\centerline{
Hirokazu Fujisaka$^1$, Hideto Shigematsu$^1$
and Bruno Eckhardt$^{2}$
}
\centerline{\it $^1$ Department of Physics, Kyushu University 33,
Fukuoka 812, Japan}
\centerline{\it $^{2}$ Fachbereich Physik der Philipps-Universit\"at,
 D-3550 Marburg,}
\centerline{\it and }
\centerline{\it  FB 8 and ICBM, C.v. Ossietzky Universit\"at,
Postfach 2503, D-2900 Oldenburg}
\centerline{\it Federal Republic of Germany}

\linespace2

\centerline{Abstract}

\noindent
A new expansion scheme to evaluate the eigenvalues of the generalized evolution
 operator (Frobenius-Perron operator) $H_{q}$ relevant to the fluctuation
 spectrum and poles of the order-$q$ power spectrum is proposed.
 The ``partition function'' is computed in terms of unstable periodic
orbits and  then used in a finite pole approximation of the continued fraction
expansion for the evolution operator. A solvable example is presented and the
approximate and exact results are compared; good agreement is found.

\linespace2


\vfill\eject

\noindent
{\bf 1. Introduction}

	The hallmark of chaos is the exponential growth of initial uncertainty,
 i.e., the sensitivity on initial condition.  This stems from the fact that all
 periodic orbits are unstable, and the representative point wanders over
the state space.  As a result the dynamical behavior of the system is quite
 irregular.  This is the fundamental reason for the necessity of a statistical
 description of chaotic systems.

	It is interesting to know, however, that despite of their instability,
 many statistical quantities are well approximated by periodic
 orbits. The periodic-orbit approximation of the natural invariant measure has
 been introduced by Sinai, Bowen, Ruelle in the 1970's [1] in the context
 of hyperbolic systems. Cycles have then been used for
 the characterization of strange sets and local expansion rates [2] and
 for the calculation of damping rates in the conventional
 double time correlation function [3].
 The main purpose of the present paper is to show how fluctuation
 and order-$q$ power spectra, which have previously been introduced
 to describe the long-time  statistics and the overall temporal correlations
 of chaotic fluctuations [4-7], can be calculated using periodic orbits and
 the continued-fraction expansion [6,8,9].
 We will thus obtain a new expansion scheme for the eigenvalues of the
 generalized Frobenius-Perron operator $H_{q}$, which plays an important role
 for the statistical description of chaotic fluctuations.

	The present paper is organized as follows.  In section 2 we give a brief review
 of our approach to dynamical fluctuations.  In section 3
 we will derive fundamental equations which determine the eigenvalues in terms
 of unstable periodic orbits;
 they generalize the ones studied in Ref.~3.
 A new scheme for the estimation of the eigenvalues is proposed in
 section 4 by using the continued-fraction expansion.  A solvable
 non-obvious example is presented in section 5, and finite-pole
 approximations of the continued-fraction expansion are compared with the exact
 results.  The paper is closed in section 6 with some final remarks.
 The relation between the generating function and the generalized evolution
 operator and the cycle expansion of the Fredholm determinant are discussed in
 appendices.

\linespace1

\noindent
{\bf 2. Statistical characterization of dynamical fluctuations}

	Consider a dissipative, chaotic $d$-dimensional map
$${\bmit x}(j+1)={\bmit f}({\bmit x}(j)),\qquad
 {\bmit x}=(x_{1},x_{2},\cdots, x_{d}).\eqno (2.1)$$
Let $u_{j}=u\{{\bmit x}(j)\}$ be an observed time series which is a unique,
scalar function of ${\bmit x}(j)$ generated by (2.1).  The
 probability distribution $p_{n}(u)$ that the average
$${\bar u}_{n}({\bmit x}(0))={1\over n}\sum^{n-1}_{j=0}u\{{\bmit x}(j)\}, \eqno
 (2.2)$$
takes a value $u$ is asymptotically given by [4,6,10]
$$\eqalign{
p_{n}(u)&\equiv <\delta({\bar u}_{n}-u)>\cr
&=\int \delta({\bar u}({\bmit x})-u)\rho_{*}({\bmit x})dx\cr
&\sim e^{-s(u)n},\cr}\eqno (2.3)$$
for large $n$.  Here $\rho_{*}({\bmit x})$ is the natural invariant
 density, satisfying
$$\rho_{*}({\bmit x})=\int\delta({\bmit x}-{\bmit f}({\bmit y}))
\rho_{*}({\bmit y})d{\bmit y}\equiv H\rho_{*}({\bmit x}), \eqno (2.4)$$
with the Frobenius-Perron (FP) operator $H$.
The generating  function
$$M_{q}(n)=<\exp[q\sum^{n-1}_{j=0}u\{{\bmit x}(j)\}]>, \eqno (2.5)$$
asymptotically takes the form
$$M_{q}(n)\sim e^{\phi(q)n}, \eqno (2.6)$$
for large $n$.  The functions $\phi$ and $s$ are related to each other
 via [4,6,10]
$$\phi(q)=-\min_{u}[s(u)-qu].\eqno(2.7)$$

	The function $\phi(q)$ plays a fundamental role for the overall
 characterization of temporal fluctuations.  However it cannot describe the
 explicit temporal correlations in $\{u\{{\bmit x}(j)\}\}$.  These
 can be singled out as follows.  Let us define a weighted average
$$\eqalign{
u(q)&=d\phi(q)/dq\cr
&=\lim_{n\to \infty}\langle{\bar u}({\bmit x}(0))
e^{qn{\bar u}_{n}({\bf x}(0))}\rangle
/M_{q}(n)\,,\cr} \eqno (2.8)$$
and the order-$q$ power spectrum [7]
$$\eqalign{
I_{q}(\omega)&=\lim_{n\to\infty}\langle F_{n}(\omega,{\bmit x}(0))\delta(
{\bar u}_{n}({\bmit x}(0))-u(q))\rangle/p_{n}(u(q))\cr
 &=\lim_{n\to\infty}\langle F_{n}(\omega,{\bmit x}(0))
e^{qn{\bar u}_{n}({\bf x}(0))}\rangle
/M_{q}(n)\,,\cr}\eqno (2.9)$$
where
$$F_{n}(\omega,{\bmit x})\equiv {1\over n}\left|\sum^{n-1}_{j=0}[
u\{{\bmit f}^{j}({\bmit x})\}-u(q)]
e^{-i\omega j}\right|^{2}\,.\eqno (2.10)$$
Then $I_{q}(\omega)$ is identical to the power spectrum over time regions where
 the coarse-grained quantities ${\bar u}_{n}$ take the same value
$u(q)$ and singles out the temporal correlation characteristic of the intensive
 parameter $q$.  In addition, one can define the generalized double time
 correlation function $C_{q}(n)$ by an ensemble average as above.
 It turns out that $I_{q}(\omega)$ and $C_{q}(n)$ are related to each other via
 the  Wiener-Khinchin theorem.

	If there exist only poles, one obtains [7]
$$I_{q}(\omega)={1\over 2}{\sum_{l}}'K^{(l)}_{q}
{\sinh[\gamma^{(l)}_{q}+i\omega^{(l)}_{q}]\over\sinh^{2}
[{\gamma^{(l)}_{q}+i\omega_{q}^{(l)}\over 2}]+\sin^{2}({\omega\over 2})}\, ,
\eqno (2.11)$$
and the corresponding order-$q$ double time correlation function takes the
 form
$$C_{q}(n)={\sum_{l}}'K^{(l)}_{q}\exp
[-\{\gamma^{(l)}_{q}+i\omega_{q}\}n]\, , \eqno (2.12)$$
where $\{\gamma^{(l)}_{q}\}$ and $\{\omega^{(l)}_{q}\}$ are respectively the
 sets of decay rates and characteristic frequencies.  The ordinary
 power
 spectrum [11] is given by $I_{q=0}(\omega)$ [7].
\linespace1

\noindent
{\bf 3. Determination of eigenvalues of $H_{q}$ by periodic cycles}

In appendix A (see also [5]), we show that eq.~(2.5) can be rewritten as
$$M_{q}(n)=\int H_{q}^{n}\rho_{*}({\bmit x})d{\bmit x}\,,\eqno (3.1)$$
where $H_{q}$ is the generalized Frobenius-Perron (GFP) operator
 defined by [5,12]
$$H_{q}F({\bmit x})=\int\delta({\bmit x}-{\bmit f}({\bmit y}))e^{qu
\{{\bf y}\}}
F({\bmit y})d{\bmit y}\,; \eqno (3.2)$$
$H_{0}$ is identical to the usual FP operator.  Let $\nu^{(l)}_{q}$
and $\psi^{(l)}_{q}$ be the eigenvalues and eigenfunctions of $H_{q}$,
$$H_{q}\psi^{(l)}_{q}({\bmit x})=\nu^{(l)}_{q}\psi^{(l)}_{q}({\bmit x})\,. \eqno
 (3.3)$$
Expanding $\rho_{*}({\bmit x})$ in terms of $\{\psi^{(l)}_{q}({\bmit x})\}$,
 and inserting it into (3.1) we obtain
$$M_{q}(n)=\sum_{l}J^{(l)}_{q}[\nu^{(l)}_{q}]^{n}\,,\eqno (3.4)$$
with expansion coefficients $\{J^{(l)}_{q}\}$, $(\sum_{l}J^{(l)}_{q}=1)$.
  Here and throughout  the paper we assume that $H_{q}$
 has only discrete eigenvalues.   Therefore, the characteristic function
$\phi(q)$ is determined by the largest eigenvalue $\nu^{(0)}_{q}$,
 which is positive because of the Perron theorem, viz.
$$\phi(q)=\log\nu^{(0)}_{q}, \eqno (3.5)$$
$(\nu^{(0)}_{0}=1)$.

	The order-$q$ power spectrum and the double time correlation function
can be expressed as in (2.11), the poles being determined by
$$\exp[\gamma^{(l)}_{q}+i\omega^{(l)}_{q}]=\nu^{(0)}_{q}/\nu^{(l)}_{q}, \quad
 (l\ne0)\,.\eqno (3.6)$$
The coefficients in (2.11) are obtained from [7]
$$\eqalign{
K^{(l)}_{q}&=a_{q}(0,l)a_{q}(l,0)\cr
u\{{\bmit x}\}\psi^{(l)}_{q}({\bmit x})&=\sum_{l'}a_{q}(l,l')\psi^{(l')}_{q}
({\bmit x})\,,\cr}\eqno (3.7)$$
assuming completeness of the eigenfunctions.  The functions
$I_{q}(\omega)$ and $C_{q}(n)$ concern the explicit temporal correlation as
 singled out with the parameter value $q$ from infinitely many aspects of
 temporal-correlation characteristics embedded in $\{{\bmit u}\{{\bmit
 x}(j)\}\}$.
  The ordinary double time correlation function
$C(n)\equiv \langle\delta u\{{\bmit x}(n)\}\delta u\{{\bmit x}(0)\}\rangle$,
 (where $\delta u=u-\langle u\rangle$), simply describes only one form,
 corresponding to the limit $q\to 0$, $(C(n)=C_{0}(n))$.

	Let us define the ${\bmit xy}$ element of the operator $H_{q}$ by
$$(H_{q})_{\bf xy}\equiv \delta({\bmit x}-{\bmit f}({\bmit y}))
e^{qu\{{\bf y}\}}\,,\eqno (3.8)$$
and therefore
$$(H_{q}^{n})_{\bf xy}=\delta({\bmit x}-{\bmit f}^{n}({\bmit y}))
\exp[q\sum^{n-1}_{j=0}u\{{\bmit f}^{j}({\bmit y})\}]\,.\eqno (3.9)$$
The partition function
$$Z_{q}(n)\equiv {\rm Tr\ }H^{n}_{q}=\int(H^{n}_{q})_{\bf xx}\,d{\bmit x}\,,
 \eqno(3.10)$$
is expanded as
$$\eqalign{
Z_{q}(n)&=\int\delta({\bmit x}-{\bmit f}^{n}({\bmit x}))\exp
[q\sum^{n-1}_{j=0}u\{{\bmit f}^{j}({\bmit x})\}]d{\bmit x}\cr
&=\sum_{k}
{\exp[q\sum^{n-1}_{j=0}u\{{\bmit f}^{j}({\bmit x}^{(n),k})\}]\over
|\det[{\hat 1}-{\hat D}_{n}({\bmit x}^{(n),k})]|}\,.\cr}
\eqno (3.11)$$
Here the summation $\sum_{k}$ extends over all fixed points
$\{{\bmit x}^{(n),k}\}$, where
 ${\bmit x}^{(n),k}$ denotes the $k$-th fixed point of the $n$-th iterate
 ${\bmit f}^{n}({\bmit x})$, $(k=1,2,\cdots,n)$,
$${\bmit x}^{(n),k}={\bmit f}^{n}({\bmit x}^{(n),k})\,.\eqno (3.12)$$
Furthermore $({\hat D}_{n}({\bmit x}))_{jk}=
\partial({\bmit f}^{n}({\bmit x}))_{j}/\partial
 x_{k}$, i.e.,
$${\hat D}_{n}({\bmit x})={\prod^{n-1}_{j=0}}{}_{+}\,{\hat g}({\bmit f}^{j}
({\bmit x}))\,,\quad ({\hat g}({\bmit x}))_{kl}={\partial f_{k}({\bmit x})\over
\partial x_{l}}\,, \eqno (3.13)$$
$\prod_{+}$ being the ordered product.

	The spectral decomposition of $H_q$ on the other hand yields
$${\rm Tr\ }H^{n}_{q}=\sum_{l}[\nu^{(l)}_{q}]^{n}\,. \eqno (3.14)$$
Together with (3.11) we thus obtain a set of fundamental relations between the
eigenvalues $\{\nu_q^{(l)}\}$ and the periodic cycles as contained in
$Z_q(n)$, viz.
$$\sum_{l}[\nu^{(l)}_{q}]^{n}=Z_{q}(n)\,,\eqno (3.15)$$
$(n=1,2,3,\cdots)$.
 The next problem is to solve for the $\{\nu^{(l)}_{q}\}$ in terms of
 $\{Z_{q}(n)\}$ which are determined by periodic orbits as in (3.11).
 In Ref.~[3] and appendix B, the cycle expansion method is used for this
 purpose. Our main aim here is to study the applicability of a continued
 fraction expansion, to which we turn now.

\linespace1

\noindent
{\bf 4. Continued-fraction expansion for the eigenvalues}

	Let $A[\mu]$ denote the Laplace transform of a function $A(n)$,
$$A[\mu]\equiv \sum^{\infty}_{n=1}A(n)\mu^{n-1}\,.\eqno (4.1)$$
Applied to eq.(3.15) one finds
$$Z_{q}[\mu]=\sum_{l}({1\over \nu^{(l)}_{q}}-\mu)^{-1}. \eqno (4.2)$$
The eigenvalues $\{\nu^{(l)}_{q}\}$ are thus determined by the poles of
 $Z_{q}[\mu]$, which we estimate from a continued fraction expansion.

     For convenience, let us put
$$(n)_{1}\equiv Z_{q}(n), \eqno (4.3)$$
$(n=1,2,3,\cdots)$.  Let $(n)_{1}$ obey the ``equation of motion'',
$$(n+1)_{1}=b_{1}(n)_{1}+\sum^{n-1}_{j=1}(n-j)_{2}(j)_{1}, \eqno (4.4)$$
where $b_{1}=(2)_{1}/(1)_{1}$.  In the memory term a new function $(n)_{2}$
 appears.  By constructing an equation of motion for
$(n)_{2}$ similarly to (4.4), a new function $(n)_{3}$ enters. Repeating
 this procedure successively, we have a set of functions
$\{(n)_{k}, k=1,2,\cdots\}$, related to each other via
$$(n+1)_{k}=b_{k}(n)_{k}+\sum^{n-1}_{j=1}(n-j)_{k+1}(j)_{k}, \eqno (4.5)$$
where $b_{k}\equiv(2)_{k}/(1)_{k}$, $(k=1,2,\cdots)$.  Equation (4.5)
 is regarded as the equation of motion for $(n)_{k}$ determined by the memory
 kernel $(n)_{k+1}$.

	The Laplace transform of $(n)_{k}$ is given by
$$\eqalign{
[\mu]_{k}&\equiv\sum^{\infty}_{n=1}(n)_{k}\mu^{n-1},\cr
&={(1)_{k}\over 1-b_{k}\mu-[\mu]_{k+1}\mu^{2}}.\cr}\eqno (4.6)$$
Thus we get the continued-fraction expansion [6]
$$Z_{q}[\mu]=
{(1)_{1}\over\displaystyle 1-b_{1}\mu-
{\mu^{2}(1)_{2}\over\displaystyle 1-b_{2}\mu-
{\mu^{2}(1)_{3}\over\displaystyle 1-b_{3}\mu-
{\mu^{2}(1)_{4}\over\displaystyle 1-b_{4}\mu-\cdots}}}}. \eqno (4.7)$$
Since $\{\nu^{(l)}_{q}\}$ are identical to the reciprocal poles of
$Z_{q}[\mu]$, we find
$$\nu^{(l)}_{q}=1/\mu^{(l)}_{q}, \eqno (4.8)$$
where $\{\mu^{(l)}_{q}\}$ are poles of (4.7), i.e., the zeros
of the denominator.
Obviously, the continued-fraction
representation (4.7) can be expanded as (4.2).

	Now let us turn to how to determine the coefficients
$\{(1)_{k},b_{k},k=2,3,4,\cdots\}$ in (4.7).  One should note that the
 quantity $(n)_{k+1}$ is determined from $\{(j)_{k},j=1,2,\cdots, n+2\}$ by
$$(n)_{k+1}={1\over (1)_{k}}\{
(n+2)_{k}-b_{k}(n+1)_{k}-\sum^{n-1}_{l=1}(l)_{k+1}(n+1-l)_{k}\}, \eqno (4.9)$$
with
$$(1)_{k+1}={1\over (1)_{k}}\{(3)_{k}-b_{k}(2)_{k}\}, \eqno (4.10)$$
$(k=1,2,\cdots)$.  So the quantity $(n)_{k}$ is given in terms
 of $\{(j)_{1},j=1,2,\cdots,n+2(k-1)\}$.  Especially, parameters $(1)_{k}$
 and $(2)_{k}$, $(k=2,3,4,\cdots)$, in (4.7) are given, respectively, by sets
 $\{Z_{q}(j),j=1,2,\cdots, 2k-1\}$ and $\{Z_{q}(j), j=1,2,\cdots, 2k\}$.
  How $(1)_{k}$ and $(2)_{k}$ are determined by $\{Z_{q}(j)\}$ is given by
solving the recursion relation (4.9) with the initial condition (4.10).

	Since from a practical viewpoint the infinite continued-fraction
expansion is intrac\-table,
we must develop a discard approximation.  This can
be carried out with the finite-pole approximation of the continued-fraction
expansion as follows.
  Let us assume that the eigenvalues are ordered,
 $\nu^{(0)}_{q}>|\nu^{(1)}_{q}|\ge|\nu^{(2)}_{q}|\ge\cdots$,
 and let us terminate (3.15) after $m$ terms, i.e.,
$$\sum^{m-1}_{l=0}[\nu^{(l)}_{q}]^{n}=Z_{q}(n). \eqno (4.11)$$
  This is reasonable, if the eigenvalues with $l\le m$
decay sufficiently rapidly.
This is equivalent to the $m$-pole
approximation $(1)_{m+1}=0,i.e.,$
$$[\mu]_{m}={(1)_{m}\over 1-b_{m}\mu}\,, \eqno (4.12)$$
and leads to
$$[\mu]_{k}={(1)_{k}\sum^{m+1-k}_{l=1}a_{k+1}(l)\mu^{l-1}\over
\sum^{m+2-k}_{l=1}a_{k}(l)\mu^{l-1}}, \quad (k=1,2,\cdots,m). \eqno (4.13)$$
Here coefficients are
$$a_{k}(1)=1, \qquad (k=1,2,\cdots,m+1), \eqno (4.14)$$
$$a_{k}(2)=-\sum^{m}_{j=k}b_{j}, \qquad (k=1,2,\cdots,m), \eqno(4.15)$$
$$a_{m-1}(3)=-b_{m-1}a_{m}(2)-(1)_{m}. \eqno (4.16)$$
Furthermore
$$a_{k}(l)=a_{k+1}(l)-b_{k}a_{k+1}(l-1)-(1)_{k+1}a_{k+2}(l-2), \eqno (4.17)$$
for $k=1,2,\cdots,m-2$ and $3$
{\vbox spread -2pt{\hbox{$<$}\hrule\vskip1pt\hrule}}
$l${\vbox spread -2pt{\hbox{$<$}\hrule\vskip1pt\hrule}}
$m-k+1$, and
 $$a_{k}(l)=-b_{k}a_{k+1}(l-1)-(1)_{k+1}a_{k+2}(l-2), \eqno (4.18)$$
for $l=m-k+2$ where $k=1,2,\cdots, m-2$.

	The poles $\{\mu^{(l)}_{q}\}$ with the $m$-pole approximation
 are therefore the roots of the order-$m$ algebraic equation
$$\Omega_{m}(\mu)\equiv \sum^{m+1}_{l=1}a_{1}(l)\mu^{l-1}=0. \eqno (4.19)$$
By putting $\nu=1/\mu$, the equation $\Omega_{m}(1/\nu)=0$ is therefore
 identical to the characteristic equation of (3.3) with the $m$-eigenstate
 approximation.

	Let us comment on the validity of the finite-pole approximation of the
 continued-fraction expansion.  It is obvious that the expansion (4.5)
 is meaningful only when $\{|b_{k}|\}$ are not too large.  Assume that for one
 $k$-value, say $k=K($
\vbox spread -2pt{\hbox{$<$}\hrule\vskip1pt\hrule}
$m)$, $(1)_{K}$ vanishes (or, in a practical sense, is extremely small) for some
 $q$-value, $q=Q$.  The quantity  $|b_{K}|$ then
 diverges at $q=Q$,  if $(2)_{K}$ remains finite.  Since a part of
 $\Omega_{m}(\mu)$ contains terms $h(\mu)b_{K}\mu$ at $q=Q$, $h$ being a
 non-singular factor,
 the algebraic equation $\Omega_{m}(\mu)=0$ has a solution $\mu=0$ at $q=Q$, i.e
 one reciprocal eigenvalue $\nu$ diverges.  As $q$ is changed from below
$Q$, one reciprocal eigenvalue gradually, say, increases, and diverges
 at $q=Q$, i.e. $\nu=\infty$ for $q=Q-0$.  For $q=Q+0$, $\nu=-\infty$.  Near
 such points the continued-fraction expansion loses its validity.

\linespace1

\font\fontA=cmtt10 scaled \magstep3
\def\bigzero{\mathrel{\hbox{\fontA 0}}}
\def\bigasterisk{\mathrel{\hbox{\fontA *}}}
\def\calF{\mathrel{\hbox{$\cal F$}}}
\def\geqq{\mathrel{\hbox{\raise0.15ex
  \hbox{$\ge$}\kern-0.77em\raise-0.45ex\hbox{$-$}}}}
\def\leqq{\mathrel{\hbox{\raise0.15ex
  \hbox{$\le$}\kern-0.77em\raise-0.45ex\hbox{$-$}}}}

\leftline {\bf 5. Example --- Exact eigenvalues and finite-pole approximation
 ---}
\par
In this section we study a simple non-obvious solvable example and
compare the finite-pole approximation of the continued-fraction
expansion with exact results.
The model we discuss is the Markov map
$$
f(x) = \cases{ s_0 x \qquad & $(0\leqq x \leqq a)$ \cr
\quad & \cr
s_1 (x-a) \qquad & $(a < x \leqq 1)$, \cr } \eqno(5.1)
$$
where $s_0 \equiv a^{-1}$ and $s_1 \equiv a/(1-a)$.

\linespace1
\par
\leftline{A. Partition function and eigenvalues}
First, we calculate the partition function $Z_q(n)$ from periodic orbits.
Let $I_0=[0,a]$ and $I_1=(a,1]$ be the dynamical partition of the full interval
$I=[0,1]$, i.e., the subintervals on which $f$ has constant slope. $f$ maps
$I_0$
onto $I$ and $I_1$ onto $I_0$. Higher iterates of $f$ give rise to finer
subdivisions of $I$ which can be labelled by strings of letters $A$ and $B$,
depending on whether the iterates come to lie in the subintervals $I_0$ or $I_1$
respectively. Since $I_1$ is mapped onto $I_0$, there can be no repetitions of
the letter $B$, i.e., every $B$ is followed by an $A$. If the last letter of
a string of length $k$ is $A$, then the $k$-th iterate of the subinterval is
mapped
onto the full interval and thus has a fixed point (cycle of period $k$). If the
letter is a $B$, then there will be a fixed point only if $x\in I_0$.

The derivatives of $f^{(k)}$ at its fixed points are $s_0^{n_A} s_1^{n_B}$,
where $n_A$ and $n_B$ are the number of intermediate
points in the intervals $I_0$ and $I_1$,
respectively. Since they depend only on the number and not on the ordering,
periodic points with strings differing only in the ordering of symbols will
give rise to the same derivatives and the same weights in the traces,
eq.~(3.11).

We now turn to the computation of the number of points with the same
derivatives.
To this end, we consider the sum of slopes, $W_T (k)$, over all fixed points
of $f^k (x)$.
A fixed point of $f^k(x)$ on a subinterval is classified into
two groups $A$ and $B$ according to the last letter of the string.
Let $W^A_{\sigma}(k)$, ($\sigma = 0$ or 1) denote the sum of
slopes over the
fixed points in the group $A$ which are on $I_{\sigma}$.  Similarly,
$W^B_{\sigma}(k)$ is defined for the fixed points in the group $B$.
Then, we can write the recursion equation
$$
\left. \eqalign{ W^A_0 (k) &= s_0 \bigl[ W^A_0 (k-1) + W^B_0 (k-1)\bigr]\cr
W^B_0 (k) &= W^A_1 (k) = s_1 W^A_0 (k-1) \cr
W^A_0 (1) &= s_0\,, \qquad W^A_0 (0) = 0\,. \cr} \right\} \eqno(5.2a)
$$
The sum $W_T (k)$ can be written as
$$
W_T (k) = W^A_0 (k) + W^B_0 (k) + W^A_1 (k)\,.
\eqno(5.2b)
$$
The recursion equation $(5.2a)$ leads to
$$
W^A_0 (k) = {s_0 \over y_+ - y_-}\Bigl[ y^k_+ - y^k_- \Bigr]\,,
\eqno(5.3)
$$
where we put
$$
y_{\pm} = {1\over 2} \Bigl[ s_0 \pm \sqrt{ s^2_0 + 4s_0s_1 }\> \Bigr]\,.
\eqno(5.4)
$$
Inserting (5.3) and (5.4) into $(5.2b)$, we have
$$
\eqalignno{ W_T (k) &= y^k_+ + y^k_-  \cr
&= \Bigl({1 \over 2}\Bigr)^{k-1} \sum_{i=0}^{[k/2]}\, {k \choose 2i}\,
\sum_{j=0}^i\, {i \choose j}\, \bigl( 4s_1\bigr)^j s_0^{k-j}\,, &(5.5) \cr}
$$
where $[\cdot]$ is  Gauss's symbol.  Because the slopes of $f^k(x)$
are written in the factorized forms $s^j_1 s^{k-j}_0\
(j=0,1,2,\ldots,[k/2])$, the number of the fixed points with the same slope
is given by the corresponding coefficient of the homogeneous equation (5.5).

If we now specify the observable $u\{x\}$ to be the characteristic function
on the interval $[a,1]$, i.e.,
$$
u\{x\} = \cases{\  0 \qquad & $(0 \leqq x \leqq a)$ \cr \noalign{\vskip 5pt}
\  1 \qquad & $(a < x \leqq 1) \,$, \cr}
\eqno(5.6)
$$
we can write the partition function (3.11) as
$$
Z_q (n) = \Bigl({1 \over 2}\Bigr)^{n-1} \sum_{i=0}^{[n/2]}\,
{n \choose 2i}\, \sum_{j=0}^i \, {i \choose j}
\bigl( 4 e^q \bigr)^j \big/ \bigl( s_1^j s_0^{n-j} - 1\bigr).
\eqno(5.7)
$$

Now we evaluate eigenvalues of the operator $H_q$ with observable (5.6).
As we are interested in an absolutely continuous measure,
we consider the functional
space $\calF$ which consists of all linear combinations of polynomials on
$I_0$ and $I_1$.  Let $\chi_{\sigma}$, $(\sigma = 0$ or $1)$ denote the
characteristic function of $I_{\sigma}$: $\chi_{\sigma}(x) = 1$ for $x \in
I_{\sigma}$, and 0 for $x \notin I_{\sigma}$.  Then, a linear combination
$G \in \calF$ is defined by
$$
G(x) = P_j(x)\chi_0(x) + P_k(x)\chi_1(x),
\eqno(5.8)
$$
where $P_j$ and $P_k$ are polynomials on $I$.  Because the map (5.1) is
piecewise linear and the $u\{x\}$ piecewise constant, the map $H_q: \calF
\rightarrow \calF$,
$$
\left( H_q G\right) (x) = \Bigl[ s_0^{-1} P_j \bigl(s_0^{-1}x\bigr) +
s_1^{-1}P_k \bigl( s_1^{-1}x + a\bigr) e^q \Bigr] \chi_0 (x) +
s_0^{-1} P_j \bigl(s_0^{-1}x \bigr) \chi_1 (x)
\, ,\eqno(5.9)
$$
preserves the maximal degree of polynominals.
For $n = 0,1,2,\ldots$, the set of linear combinations of polynomials of
degree $k$ $(0 \leqq k \leqq n)$ is a subspace of $\calF$, denoted by
$\calF^{(n)}$.  The restriction of $H_q$ onto $\calF^{(n)}$ determines an
operator $H_q^{(n)} :\calF^{(n)} \rightarrow \calF^{(n)}$.  As a linear
combination $G(x)$ in $\calF^{(n)}$ can be represented by a
$2(n+1)$-dimensional vector [11], and $H_q^{(n)}$ can be written as a $2(n+1)
\times 2(n+1)$ matrix.  Indeed, we have the upper-triangular matrix
\def\matrixA{\mathrel{\hbox{$\matrix{h^{(0)}_q & \cr & h^{(1)}_q \cr}$}}}
\def\matrixB{\mathrel{\hbox{$\matrix{\ddots & \cr & h^{(n)}_q \cr}$}}}
$$
H^{(n)}_q = \left( \matrix{\matrixA &\bigasterisk \cr
\raise-0.75ex\hbox{$\bigzero$} \quad &\matrixB \cr} \right),
\eqno(5.10)
$$
where each element of $H^{(n)}_q$ is a $2\times 2$ matrix and
$$
h^{(j-1)}_q \equiv \left( \matrix{ \raise-0.4ex\hbox{$s^{-j}_0$}\quad
& \raise-0.4ex\hbox{$s^{-j}_1 e^q$} \cr \quad & \quad \cr
s^{-j}_0 \quad & 0 \cr} \right),\qquad \quad(j = 1,2,\ldots,n+1).
\eqno(5.11)
$$
Eigenvalues of $h^{(j-1)}_q (1 \leqq j \leqq n+1)$ are also
eigenvalues of $H^{(n)}_q$, and are given by
$$
t^{(j-1)}_{\pm} \equiv {1\over 2}\Bigl[ s^{-j}_0 \pm
\sqrt{ s^{-2j}_0 + 4e^q (s_0 s_1)^{-j}}\, \Bigr].
\eqno(5.12)
$$
Since the full generalized Frobenius-Perron operator $H_q$ is obtained in
the limit $n \to \infty$, the eigenvalues of $H_q$ are
$$
\nu^{(l)}_q = \cases{ t^{(j)}_+ \qquad & for $l \equiv 2j = 0,2,4,\ldots$\cr
\quad & \quad \cr
t^{(j)}_- \qquad & for $l \equiv 2j + 1 = 1,3,5,\ldots$. \cr}
\eqno(5.13)
$$
Substituting (5.12) into (3.14) and summing up over $l$ yields
$$
\hbox{\rm Tr} H^n_q = \Bigl({1 \over 2}\Bigr)^{n-1} \sum_{i=0}^{[n/2]}\,
{n \choose 2i}\, \sum_{j=0}^i \, {i \choose j}
\bigl( 4 e^q \bigr)^j \big/ \bigl( s_1^j s_0^{n-j} - 1\bigr).
\eqno(5.14)
$$
We remark, that for this example the generating function (2.5)
can be expressed in terms of eigenvalues of $H^{(0)}_q$ only as
$$
M_q (n) = \Bigl\{ \bigl[ 1 - \nu^{(1)}_q \bigr]^2 \bigl[ \nu^{(0)}_q
\bigr]^n - \bigl[ 1 - \nu^{(0)}_q \bigr]^2 \bigl[ \nu^{(1)}_q \bigr]^n
\Bigr\} \Big/ \bigl[ \nu^{(0)}_q - \nu^{(1)}_q \bigr] \bigl[ 2 -
\nu^{(0)}_q - \nu^{(1)}_q \bigr].
\eqno(5.15)$$

\linespace1

\noindent
B. Truncation of the continued-fraction expansion

	We will study the approximate methods to obtain eigenvalues of the
 GFP operator for the model (5.1) by truncating the continued-fraction
 expansion.  Hereafter we consider three approximations
 with $4$-, $6$-, and $8$-poles,  for which we need traces up to  order
$n=8$, $10$ and $12$, respectively.  The parameter is chosen as $a=0.75$.

	Figure 1(a) shows  the first twenty of the exact eigenvalues (5.13).
 Figures 1(b-d) show the comparison between results of the finite-pole
 approximations and exact results.  In each approximation there exist
 apparent deviations of approximate values from
 the exact results in certain regions of $q$.
  Such deviations are located near $q$-values where one or several
 coefficients $(1)_{K}$ vanish, thus leading to a divergence of
 $|b_k|$ and a failure of the finite pole approximation of the
 continued fraction expansion.
 For the $4$-pole approximation the region showing
 apparent deviation is rather wide.  However, as the number of poles is
 increased, the number of such singular points increases, but
 the total area of effected $q$-regions
 diminishes. In this  way
 the approximate results approach the exact ones
 as the number of poles is increased.

\linespace1

\noindent
{\bf 6. Concluding remarks}

 In closing this paper let us add several remarks.  The first point concerns
 the relation between the characteristic function $\phi(q)$, especially
 for local expansion rates [13], and the decay rates $\{\gamma^{(l)}_{q}\}$ in
 $I_{q}(\omega)$.   Whereas $\phi(q)$ describes the probability
 distribution of fluctuations,
 the decay rates $\{\gamma^{(l)}_{q}\}$ characterize the
 explicit temporal correlations.  $\phi(q)$ is determined by the
 largest eigenvalue of $H_{q}$ and $\{\gamma^{(l)}_{q}\}$ by others [7].  In
 this sense, we conclude that {\it decay rates generically have no relation with
 the  characteristic function} $\phi(q)$, except in special cases.

	We proposed a continued-fraction expansion for the eigenvalues of the
 generalized Frobenius-Perron operator $H_{q}$, whose expansion coefficients are
 determined in terms of periodic orbits of the dynamical equation.  A solvable
 illustrative example was presented, and their eigenvalues were compared with
 those of truncated
 continued-fraction expansions.  It should be noted that whereas
 the largest eigenvalue is insensitive to the approximation (except in several
 narrow regions of $q$, related to the divergence of the coefficients
$\{b_{k}\}$), other eigenvalues which are relevant to the temporal correlations
 depent rather sensitively on the order of truncation in the continued
 fraction expansion. Except for this, the finite pole approximation
 of the  continued-fraction expansion turned out to be a powerful approximation.

	The ordinary double time correlation function
is determined by the eigenvalues of the FP operator
 $H$ [14].  The equations corresponding to
 (3.15) are
$${\sum_{l}}^{\prime}[\nu^{(l)}_{0}]^{n}=Z_{0}(n)-1, \eqno (6.1)$$
where $\sum'_{l}$ stands for the summation except $l=0$,
 because as a result of the existence of the invariant density the largest
 eigenvalue of $\{\nu_{0}^{(l)}\}$ is unity.
 As can be easily verified in our example, the eigenvalue
$\nu_q^{(0)} = t_+^{(0)}$ (eq.~(5.13)) indeed passes through $1$ for
$q\rightarrow 0$. In Ref.~3, eq~(6.1) was solved approximately using the
 cycle expansion of Ruelle zeta functions and Fredholm determinants
 $d_{F}$. Because of possible poles in the Ruelle zeta functions,
 one has to use the full Fredholm determinant to determine the eigenvalues.
 As the calculations in appendix B show, in our solvable example,
 the Ruelle zeta functions are entire and the Fredholm determinant factorizes
 into a product of Ruelle zeta functions; thus there are no poles and the
 spectrum as computed from Ruelle zeta functions and the Fredholm determinant
 coincides. This exact factorization of course is not noticable from a
 truncated cycle expansion of the Fredholm determinant as used in
 numerical calculations. The numerically determined eigenvalues behave very much
 as in the case of the continued fraction expansion, except that they show no
 poles.

 One might be lead to conclude from this example that the cycle expansion
 of appendix B seems to be better, but one can also find examples where
 the cycle expansion method has  problems but where the continued fraction
 expansion works fine, e.g. in the case of the fully developped parabola.
 Thus the advantages and disadvantages of the two methods are more subtle,
 and certainly require further work.

 	As reported in [7,8], $\{\nu^{(l)}_{q}\}$ can be derived as
 the poles of the Laplace transform
 of $M_{q}(n)$, where we need the explicit expression of moments $M_{q}(1)$,
 $M_{q}(2)$, $M_{q}(3)$, $\cdots$, using the explicit information on the natural
 invariant density itself.  The present approach however does not need
the natural
 invariant density to obtain $\{\nu^{(l)}_{q}\}$.

	Finally let us comment on how the order of truncation $m$ should
 be chosen.  Let $\tau$ and $\gamma$ be the characteristic periodicity and the
 damping rate, respectively.  Depending on the type of chaotic transitions
 and on the observables, $\tau$ and $\gamma$ sometimes exhibit critical
 slowing-down.  For example, for $x_{\mu}(j)$, one component of the orbit
 ${\bmit x}(t)$ itself, the damping rate $\gamma$ tends to zero with finite
 $\tau$ as the control parameter approaches the band splitting point.  And we
 find $\gamma\to 0$ and $\tau\to\infty$ near the type I intermittency.
 Although there
 are two characteristic time-scales $\tau$ and $\gamma^{-1}$, it is expected
 that $m$ should be chosen such that $m\gg\tau$, since $\tau$ is relevant to
 periodicity.  (See Ref.[15].)

\vfill\eject

\linespace1

\noindent
{\bf Appendix A --- Derivation of (3.1) ---}

	Equation (3.1) can be obtained by noting that
$$\eqalign{
M_{q}(n)&=\int\rho_{*}({\bmit x})\exp[q\sum^{n-1}_{j=0} u
 \{{\bmit f}^{j}({\bmit x})\}]d{\bmit x}\cr
&=\int\rho_{*}({\bmit x})e^{qu\{{\bf x}\}} L
 \exp[q\sum^{n-2}_{j=0}u\{{\bmit f}^{j}({\bmit x})\}]d{\bmit x},\cr}
 \eqno (A.1)$$
where $L$ is the time evolution operator,
$LG({\bmit x})=G({\bmit f}({\bmit x}))$.
Since $\int G_{1}({\bmit x})LG_{2}({\bmit x})d{\bmit x}=\int[HG_{1}({\bmit x})]
G_{2}({\bmit x})d{\bmit x}$, eq.~(A.1) can be written as
$$\int[H_{q}\rho_{*}({\bmit x})]
\exp[q\sum^{n-2}_{j=0}u\{{\bmit f}^{j}({\bmit x})\}]d{\bmit x}, \eqno (A.2)$$
where $H_{q}$ is the generalized Frobenius-Perron operator defined in (3.2).

	By repeating the above procedure successively, $M_{q}(n)$ is written as
$$\eqalign{
M_{q}(n)&=\int[H_{q}^{2}\rho_{*}({\bmit x})]\exp
[q\sum^{n-3}_{j=0}u\{{\bmit f}^{j}({\bmit x})\}]d{\bmit x}\cr
&\qquad \vdots\cr
&=\int[H_{q}^{n-1}\rho_{*}({\bmit x})]\exp
[qu\{{\bmit x}\}]d{\bmit x}.\cr} \eqno(A.3)$$
By noting $\int G({\bmit x})d{\bmit x}=\int HG({\bmit x})d{\bmit x}$, eq.(A.3)
 leads to eq.(3.1).

\linespace1

\noindent
{\bf Appendix B --- Cycle expansion and dynamical zeta function ---}

The cycle expansion approach to solving the moment problem (3.15) consists
again in forming the Laplace transform (4.1) so as to obtain the eigenvalues
as inverses of the positions of poles. However, this Laplace transform is
then expressed as a logarithmic derivative of a Zeta function, so that the
eigenvalues are determined by zeros of this Zeta function.

Consider again eq.~(3.11), expressing the trace $Z_q(n)$ in terms
of periodic orbits. Let $p$ label all primitive cycles of period $n_p$
($n_p$ is the smallest number of iterations after which the orbit closes).
Note that both
$$
u_p = \sum_{j=1}^{n-1} u\{f^j(x^{(n),k})\} \,
\eqno(B.1) $$
and
$$
\Lambda_p = |D_{n}({\bmit x}^{(n),k})|
\eqno (B.2) $$
are independent of the point where they are computed.
A cycle of period $n_p$ has $n_p$ different periodic points.
In the sum (3.11), it contributes whenever $k=n_p$ or a multiple
thereof. Thus when summing on all periods in the Laplace transform
(4.11), viz.
$$
Z_q[\mu] = \sum_{n=1}^\infty Z_q(n) \mu^{n-1} \,
\eqno (B.3) $$
one can also express this as a sum on all primitive
periodic orbits and their multiple traversals,
$$
Z_q[\mu] = \sum_p \sum_{r=1}^\infty n_p {\exp[q\, r\, u_p]
\over |1-\Lambda_p^r|} \mu^{r\, n_p-1}.
\eqno(B.4)$$
The usual manipulations (expansion of the denominator in a geometric
series and summation on $r$) [3] result in
$$
Z_q[\mu] = -\left({\partial \over \partial \mu}\right)\log
\Omega(\mu)
\eqno(B.5)$$
where the Zeta function
$$
\Omega(\mu) =
\prod_p \prod_{j=0}^\infty \left(1-\mu^{n_p} \exp(q u_p) |\Lambda_p|^{-1}
\Lambda_p^{-j}\right) = \prod_{j=0}^\infty \zeta_j^{-1} \, ,
\eqno(B.6)$$
factorizes into dynamical zeta functions $\zeta_j^{-1}$.
As we will see, for the particular map studied in section 5, each dynamical
zeta function will be a quadratic polynominal and thus contribute two
eigenvalues for each $j$.

The key to the evaluation of the products on periodic orbits is the
topological organization of the periodic orbits. In the example of
section 5, there can be no two consequtive $B$'s, since every point in the
interval $[a,1]$ is mapped back into the interval $[0,a]$. This is the
only exclusion rule. Defininig a new alphabet $0\equiv A$ and $1\equiv BA$,
one finds that all strings of $0$ and $1$ can be realized. Since moreover
the maps is linear, all weights $\Lambda_p$ factorize according to
$$
\Lambda_p = \Lambda_A^{n_0} \Lambda_{AB}^{n_1}
\eqno(B.7)$$
where $n_0$ and $n_1$ are the numbers of $0$'s and $1$'s in the symbol string
of $p$, respectively. The instabilities $\Lambda_p$ are
$$
\Lambda_A = s_0 \, ,\qquad \Lambda_{AB} = s_0 s_1 \, ,
\eqno(B.8a)$$
and the observable (5.6) becomes
$$
u_p = n_1 \, .
\eqno(B.8b)$$
One can then show [3,16] that the dynamical zeta functions become
$$
\zeta_j^{-1} = 1 - \mu |s_0|^{-1} s_0^{-j}  - \mu^2 \exp[q] |s_0s_1|^{-1}
(s_0 s_1)^{-j} \, .
\eqno(B.9)$$
which is precisely the characteristic polynominal for the submatrices
(5.11), $\zeta_j^{-1} = \det(1-\mu h_q^{(j)})$ for $j=0,1,\cdots$.
Thus in this case the cycle expansion yields the exact eigenvalues.

\vfill\eject

\linespace1

\noindent
{\bf References}

\item{[1]} See, e.g.,\hfill\break
Sinai, Ya. G.: Uspekhi Math. Nauk {\bf 27}, 27 (1972)\hfill\break
Bowen, R.: {\it Equilibrium States and the Ergodic Theory of Anosov
 Diffeomorphisms}, Lecture Notes in Mathematics {\bf 470} (New York:
 Springer-Verlag, 1975)\hfill\break
Ruelle, D.:  {\it Thermodynamic Fromalism} in Encyclopedia of Mathematics and
 Its Applications, Vol.5, (Addison-Wesley, Reading, 1978)
\item{[2]} See, e.g., the following articles:\hfill\break
Kai, T., Tomita, K.: Prog. Theor. Phys. {\bf 64}, 1532 (1980)\hfill\break
Takahashi, Y., Oono, Y.: Prog. Theor. Phys. {\bf 71}, 851 (1984)\hfill\break
Auerbach, D., Cvitanovic, P., Eckmann, J.-P., Gunaratne, G.,
 Procaccia, I.: Phys. Rev. Lett. {\bf 58}, 2387 (1987)\hfill\break
Morita, T., Hata, H., Mori, H., Horita, T., Tomita, K.: Prog. Theor. Phys. {\bf
  296 (1988)\hfill\break
Cvitanovic, P.: Phys. Rev. Lett. {\bf 61}, 2729 (1988)\hfill\break
Cvitanovic, P., Gunaratne, G. H., Procaccia, I.: Phys. Rev. A {\bf 38}, 1503
(1988)
\item{[3]}
Cvitanovi{\'c}, P., Eckhardt, B.: J. Phys. A: Math. Gen. {\bf 24}, L237 (1991)
 \hfill\break
Christiansen, F., Paladin, G., Rugh, H. H.: Phys. Rev. Lett. {\bf 65},
 2087 (1990)\hfill\break
Rugh, H. H.: Nonlinearity {\bf 5}, 1237 (1992)\hfill\break
Eckhardt, B.: Acta Phys. Pol. B {\bf 24}, (1993)  at press
\item{[4]} Fujisaka, H., Inoue, M.: Prog. Theor. Phys. {\bf 77}, 1334 (1987)
\item{[5]} Fujisaka, H., Inoue, M.: Prog. Theor. Phys. {\bf 78}, 268 (1987)
\item{[6]} Fujisaka, H.: in {\it From Phase Transitions to Chaos (Topics
 in Modern Statistical \hfill\break
Physics)}, eds.  Gy\"orgyi, G.,  Kondor, I.,
  Sasv\'ari, L., and T\'el, T., (World Scientific, 1992)
\item{[7]} Fujisaka, H., Shibata, H.: Prog. Theor. Phys. {\bf 85},
 187 (1991)\hfill\break
Just, W., Fujisaka, H.: Physica D (1993), at press
\item{[8]} Fujisaka, H., Inoue, M.: Prog. Theor.Phys. {\bf 78}, 1203 (1987)
\item{[9]} Just, W.: J. Stat. Phys. {\bf 67}, 271 (1992)
\item{[10]} Ellis, R. S.: {\it Entropy, Large Deviations, and
 Statistical Mechanics}, \hfill\break
(New York: Springer-Verlag, 1985)
\item{[11]} See, e.g., the following articles:\hfill\break
Grossmann, S., Thomae, S.: Z. Naturforsch. {\bf 32a}, 1353 (1977)\hfill\break
Mori, H., So, B. C., Ose, T.: Prog. Theor. Phys. {\bf 66}, 1266
(1981)\hfill\break
Shigematsu, H., Mori, H., Yoshida, T., Okamoto, H.: J. Stat. Phys.
{\bf 30}, 649 (1983)
\item{[12]} Sz\'epfalusy, P., T\'el, T.: Phys. Rev. A {\bf 34}, 387 (1986)
\item{[13]} Badii, R., Heinzelmann, K., Meier, P. F., Politi, A.:
 Phys. Rev. A  {\bf 37}, 1323 (1988)
\item{[14]} Mori, H., {\it et al.} and Shigematsu, {\it et al.} in Ref. [11]
\item{[15]} Just, W., {\it et al.} in Ref.[7]\hfill\break
Kobayashi, T., Fujisaka, H., Just, W.: Phys. Rev. E, (1993) at press\hfill\break
Just, W., {\it et al.}, to be submitted
\item{[16]} Artuso, R., Aurell, E., Cvitanovi{\'c}, P.: Nonlinearity
{\bf 3}, 325 and 361 (1990)\hfill\break
Cvitanovi{\'c}, P., Eckhardt, B.: Nonlinearity, (1993), at press

\linespace1
\vfill\eject
\noindent
{\bf Figure Caption}

\noindent
 Fig.1
 (a) Exact eigenvalues of $H_{q}$ for the model (5.1) with $a=0.75$.  Twenty
 eigenvalues vs $q$ are shown in the order of increasing values.
 The results from finite-pole approximations
 of the continued-fraction expansion are shown in figures
 (b) using $4$ poles, (c) using $6$ poles and (d) using $8$ poles.
  Solid and dotted lines respectively indicate results of finite-pole
 approximation and exact results.  As the number of poles is increased, the
 results tend  to agree with the exact results.

\end